\documentclass{article}
\usepackage{arxiv}
\usepackage{amsmath}
\usepackage{caption}
\usepackage{subcaption}
\usepackage[utf8]{inputenc} 
\usepackage[T1]{fontenc}    
\usepackage{hyperref}       
\usepackage{url}            
\usepackage{booktabs}       
\usepackage{amsfonts}       
\usepackage{nicefrac}       
\usepackage{microtype}      
\usepackage{lipsum}		
\usepackage{graphicx}
\usepackage{natbib}
\usepackage{doi}

\title{Fourier-domain transfer entropy spectrum}
\author{ 
Yang Tian \\
	Department of Psychology \& Tsinghua Laboratory of Brain and Intelligence\\ Tsinghua University\\ Beijing, 100084, China.\\
	\texttt{tiany20@mails.tsinghua.edu.cn} \\
	\And
	Yaoyuan Wang \\
	Data Center Technology Laboratory, Central Research Institute, 2012 Laboratories\\ Huawei Technologies Co. Ltd.\\ Beijing, 100084, China.\\
	\texttt{wangyaoyuan@huawei.com} \\
	\And
	Ziyang Zhang \\
	Data Center Technology Laboratory, Central Research Institute, 2012 Laboratories\\ Huawei Technologies Co. Ltd.\\ Beijing, 100084, China.\\
	\texttt{zhangziyang11@huawei.com} \\
	\And
	Pei Sun \\
	Department of Psychology \& Tsinghua Laboratory of Brain and Intelligence\\ Tsinghua University\\ Beijing, 100084, China.\\
	\texttt{peisun@tsinghua.edu.cn}
}

\renewcommand{\shorttitle}{Fourier-domain transfer entropy spectrum}

\begin{document}
\maketitle

\begin{abstract}
We propose the Fourier-domain transfer entropy spectrum, a novel generalization of transfer entropy, as a model-free metric of causality. For arbitrary systems, this approach systematically quantifies the causality among their different system components rather than merely analyze systems as entireties. The generated spectrum offers a rich-information representation of time-varying latent causal relations, efficiently dealing with non-stationary processes and high-dimensional conditions. We demonstrate its validity in the aspects of parameter dependence, statistic significance test, and sensibility. An open-source multi-platform implementation of this metric is developed and computationally applied on neuroscience data sets and diffusively coupled logistic oscillators.
\end{abstract}

\section{Introduction}
Measuring causality between systems is critical in numerous scientific disciplines, ranging from physics \cite{pikovsky2003synchronization} to neuroscience \cite{pereda2005nonlinear}. The challenge of such measurement arises from a widespread situation where only a limited set of time series of systems are given, and the prior knowledge of their coupling relations remains absent. Over last decades, identifying and quantifying causal relations with insufficient information has become one of the frontier problems in physics, mathematics, and statistics \cite{hlavavckova2007causality}. Substantial progress has been accomplished in measuring causality from statistics-theoretical (e.g., Granger causality \cite{granger1969investigating} and its generalizations \cite{marinazzo2008kernel,ancona2004radial,ho2004phase,arnold2007temporal,liu2009nonparanormal,dhamala2008estimating}) and information-theoretical (e.g., transfer entropy \cite{schreiber2000measuring} and its generalizations \cite{staniek2008symbolic,lobier2014phase,lungarella2007information,runge2012escaping,wollstadt2014efficient,porfiri2019transfer,restrepo2020transfer,zhang2019itene,silini2021fast}) perspectives. Although the nature of causality remains controversial \cite{pearl2000models}, these two perspectives have been demonstrated to have non-parametric connections in mathematics \cite{diks2005note,diks2001general} and share equivalent ideas in potential causality detection. Following their ideas, the existence of causality between systems $\mathsf{X}$ and $\mathsf{Y}$ (e.g., $\mathsf{X}$ is the cause of $\mathsf{Y}$) can be interpreted as 
\begin{align}
\mathcal{P}\left(\mathsf{Y}\left(t\right),\mathsf{X}_{\left[t,\beta,\Delta t\right]}\mid\mathsf{Y}_{\left[t,\beta,\Delta t\right]}\right)\neq\mathcal{P}\left(\mathsf{Y}\left(t\right)\mid\mathsf{Y}_{\left[t,\beta,\Delta t\right]}\right)\mathcal{P}\left(\mathsf{X}_{\left[t,\beta,\Delta t\right]}\mid\mathsf{Y}_{\left[t,\beta,\Delta t\right]}\right), \label{EQ1}
\end{align}
where $\mathcal{P}\left(\cdot\right)$ denotes the probability. We mark $\mathsf{X}_{\left[t,\beta,\Delta t\right]}=\left(\mathsf{X}\left(t-\beta\Delta t\right),\ldots,\mathsf{X}\left(t-\Delta t\right)\right)$ as the history of $\mathsf{X}$ with time lag unit $\Delta t$ and maximum lag number $\beta$, representing the information of system $\mathsf{X}$ during a time interval $\left[t-\beta\Delta t,t-\Delta t\right]$. In general, inequality (\ref{EQ1}) means that the uncertainty of $\mathsf{Y}$ at moment $t$ given its own historical information (term $\mathsf{Y}_{\left[t,\beta,\Delta t\right]}$) will be regulated if the historical information of $\mathsf{X}$ (term $\mathsf{X}_{\left[t,\beta,\Delta t\right]}$) is added.

Among various causality metrics, transfer entropy \cite{schreiber2000measuring} features plentiful applications (e.g., in neuroscience \cite{vicente2011transfer,victor2002binless,vakorin2010exploring,spinney2017transfer,ursino2020transfer} and economics \cite{dimpfl2013using,papana2016detecting,toriumi2018investment,camacho2021symbolic,ji2018information}) because of its robust capacity to capture non-linear causality and intrinsic connections with dynamics theory and information theory \cite{schreiber2000measuring,hlavavckova2007causality}. The initial version of transfer entropy can be defined as 
\begin{align}
\mathcal{T}\left(\mathsf{X},\mathsf{Y}\right)=\mathbb{E}_{t\in\left[\tau+1,n\right]}\Big\{\mathcal{D}\Big[\mathcal{P}\left(\mathsf{Y}\left(t\right),\mathsf{X}_{\left[t,\beta,\Delta t\right]}\mid\mathsf{Y}_{\left[t,\beta,\Delta t\right]}\right)\Big\Vert\mathcal{P}\left(\mathsf{Y}\left(t\right)\mid\mathsf{Y}_{\left[t,\beta,\Delta t\right]}\right)\otimes\mathcal{P}\left(\mathsf{X}_{\left[t,\beta,\Delta t\right]}\mid\mathsf{Y}_{\left[t,\beta,\Delta t\right]}\right)\Big]\Big\},\label{EQ2}
\end{align}
 actually equivalent to conditional mutual information \cite{cover1999elements,hlavavckova2007causality}. Equality (\ref{EQ2}) quantifies the regulating effects in inequality (\ref{EQ1}) in terms of the expectation of the mutual information between $\mathsf{Y}\left(t\right)$ and $\mathsf{X}_{\left[t,\beta,\Delta t\right]}$ given $\mathsf{Y}_{\left[t,\beta,\Delta t\right]}$. While the model-free property of transfer entropy suggests its general applicability, its application in many real situations is still limited by various deficiencies, such as dimensionality curse \cite{runge2012escaping} and noise sensitivity \cite{smirnov2013spurious}. Although extensive efforts have been devoted to optimizing transfer entropy estimation in real data (e.g., through symbolization \cite{staniek2008symbolic}, phase time-series \cite{lobier2014phase}, ensemble evaluation \cite{wollstadt2014efficient}, Lempel-Ziv complexity \cite{restrepo2020transfer}, artificial neural networks \cite{zhang2019itene}, and $k$-nearest-neighbor \cite{kraskov2004estimating}), several perspectives remain to be improved:
 \begin{itemize}
     \item[\textbf{(I)} ] Previous works primarily analyze systems as entireties, yet it is more ubiquitous that causality only originates from specific system components (e.g., there only exists causality between the low frequency components of $\mathsf{X}$ and $\mathsf{Y}$). Such causality may be veiled if we do not distinguish between different system components;
     \item[\textbf{(II)} ] Causality is time-varying. Although the initial version of transfer entropy can be calculated accumulatively or in sliding time windows, a more natural and accurate measurement remains absent;
     \item[\textbf{(III)} ] Probability density estimation, a prerequisite of transfer entropy calculation, may be costly (e.g., critically requires a large sample size) or inaccurate (e.g., in non-stationary systems and high-dimensional spaces). Although some approaches (e.g., ensemble evaluation in cyclo-stationary systems \cite{wollstadt2014efficient,gardner2006cyclostationarity} and $k$-nearest-neighbor estimation \cite{kraskov2004estimating} in high-dimensional spaces) are developed to optimize the estimation, a low-cost transfer entropy generalization born to deal with non-stationary systems is demanded. 
 \end{itemize}
 
 In this letter, we discuss the possibility of generalizing transfer entropy into a form that maintains its original advantages and makes progress in aspects of \textbf{(I-III)}. To devote to open science, we release a multi-platform code implementation of our metric and demonstrate its validity on multiple data sets in neuroscience and physics.
 
 \section{Fourier-domain representation of systems}
An arbitrary system $\mathsf{X}$ can be subdivided into a set of components according to different criteria. To maintain the model-free property of transfer entropy measurement, we concentrate on a universal criterion subdividing $\mathsf{X}$ into different frequency components, namely Fourier-domain (frequency-domain) representation. Such a representation features widespread applications in physics \cite{marks2009handbook}, mathematics \cite{bremaud2013mathematical}, neuroscience \cite{bruns2004fourier}, and engineering \cite{rabiner1975theory}, partly ensuring the general applicability of our approach. 
 
 To avoid the stationary assumption, one can derive the Fourier-domain representation $\mathcal{F}\left(\mathsf{X},t,\omega\right)$ of $\mathsf{X}$ (here $\omega$ denotes frequency) applying wavelet (WT) \cite{rioul1991wavelets} or Hilbert-Huang transform (HHT) \cite{huang2014hilbert} rather than Fourier transform (FT) owing to the limitations of FT for non-stationary processes \cite{kaiser1994friendly}. In general, WT convolves $\mathsf{X}$ with oscillatory kernels of different frequency bands to realize time-frequency localization \cite{rioul1991wavelets} while HHT directly measures the instantaneous frequency of $\mathsf{X}$ \cite{huang2014hilbert}. Although WT and HHT support time-frequency analysis irrespective of stationary assumption, they are still not ideal owing to certain limitations and should be selected according to circumstances \cite{folland1997uncertainty,huang2014hilbert}. Here we choose WT to conduct our derivations owing to its complete mathematical foundations, yet other frameworks (e.g., HHT and its variant \cite{peng2005improved}) can also be applied. 
 
In terms of continuous WT spectrum, the Fourier-domain representation $\mathcal{F}\left(\mathsf{X},t,\omega\right)$ can be obtained through \cite{ogden1997essential,chun2010tutorial,coifman1992wavelet}
 \begin{align}
 \mathcal{F}\left(\mathsf{X},t,\omega\right)&=\mathsf{W}_{\psi}\big(\mathsf{X},t,\zeta_{\psi}\left(\omega\right)\big)\mathsf{W}_{\psi}^{\heartsuit}\big(\mathsf{X},t,\zeta_{\psi}\left(\omega\right)\big),\label{EQ3}\\
 \mathsf{W}_{\psi}\left(\mathsf{X},t,s\right)&=\frac{1}{\sqrt{s}}\int\psi^{\heartsuit}\left(\frac{\tau-t}{s}\right)\mathsf{X}\left(\tau\right)d\tau,\label{EQ4}
 \end{align}
 where $\psi\in L^{2}\left(\mathbb{R}\right)$ denotes a mother wavelet function (e.g., Morlet wavelets) and notion $\heartsuit$ represents conjugate. Notions $s$ and $\tau$ denote the scaling (frequency localization) and the translation (time localization) parameters, respectively. Function $\zeta_{\psi}\left(\cdot\right)$ is the bijective mapping from frequency $\omega$ to scale $s$, varying according to the selection of mother wavelet function $\psi$. Because the approaches in (\ref{EQ3}-\ref{EQ4}) have been extensively studied and applied \cite{ogden1997essential,chun2010tutorial,coifman1992wavelet}, we do not repeat their derivations.

 The Fourier-domain representation $\mathcal{F}\left(\mathsf{X},t,\omega\right)$ can function as a time-frequency spectrum $\left(t,\omega\right)$ of system $\mathsf{X}$, where spectrum value denotes the power. In Fig. \ref{G1a}, we show the Fourier-domain representation of an open-source near infrared spectroscopy (NIRS, a kind of electromagnetic-spectrum-based brain imaging approach) data set recorded in the superior frontal cortex \cite{cui2012nirs,dataset}. Function $\psi$ is set as the Morlet wavelet. Apart from $\mathcal{F}\left(\mathsf{X},t,\omega\right)$ (subject $1$) and $\mathcal{F}\left(\mathsf{Y},t,\omega\right)$ (subject $2$), we also present $\mathcal{F}\left(\mathsf{XY},t,\omega\right)=\mathsf{W}_{\psi}\big(\mathsf{X},t,\zeta_{\psi}\left(\omega\right)\big)\mathsf{W}_{\psi}^{\heartsuit}\big(\mathsf{Y},t,\zeta_{\psi}\left(\omega\right)\big)$.
 
  \section{Fourier-domain transfer entropy spectrum}
Once $\mathcal{F}\left(\mathsf{X},t,\omega\right)$ and $\mathcal{F}\left(\mathsf{Y},t,\omega\right)$ have been given, we turn to quantify the time-varying latent causality between them in terms of transfer entropy. Below, we discuss one possibility of this measurement.
 
\begin{figure}
     \centering
     \begin{subfigure}[b]{0.48\columnwidth}
         \includegraphics[width=\columnwidth]{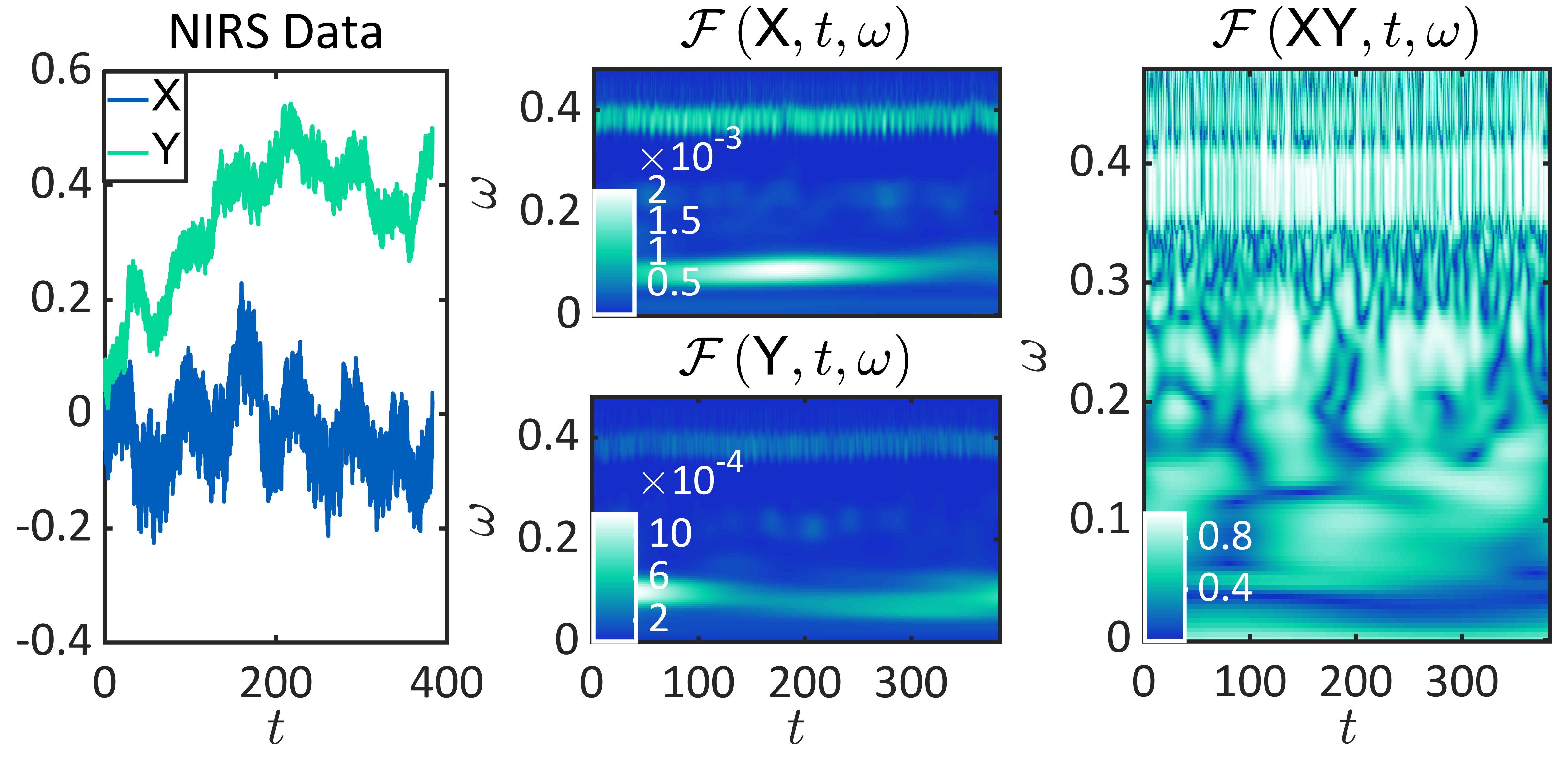}
        \caption{\label{G1a}}
     \end{subfigure}
     \hfill
     \begin{subfigure}[b]{0.48\columnwidth}
         \includegraphics[width=\columnwidth]{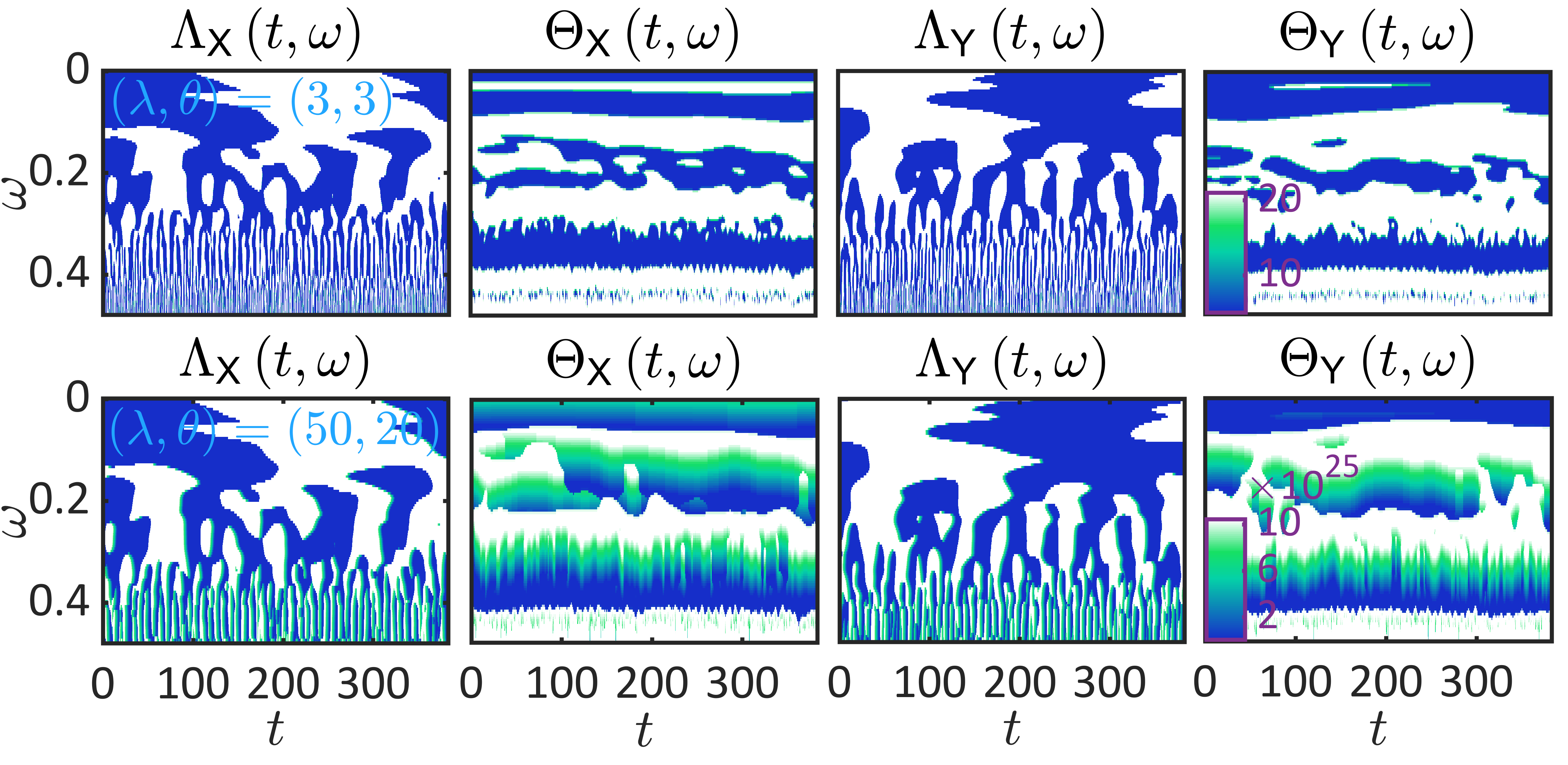}
     \caption{\label{G1b}}
     \end{subfigure}
     \hfill
     \begin{subfigure}[b]{0.48\columnwidth}
         \includegraphics[width=\columnwidth]{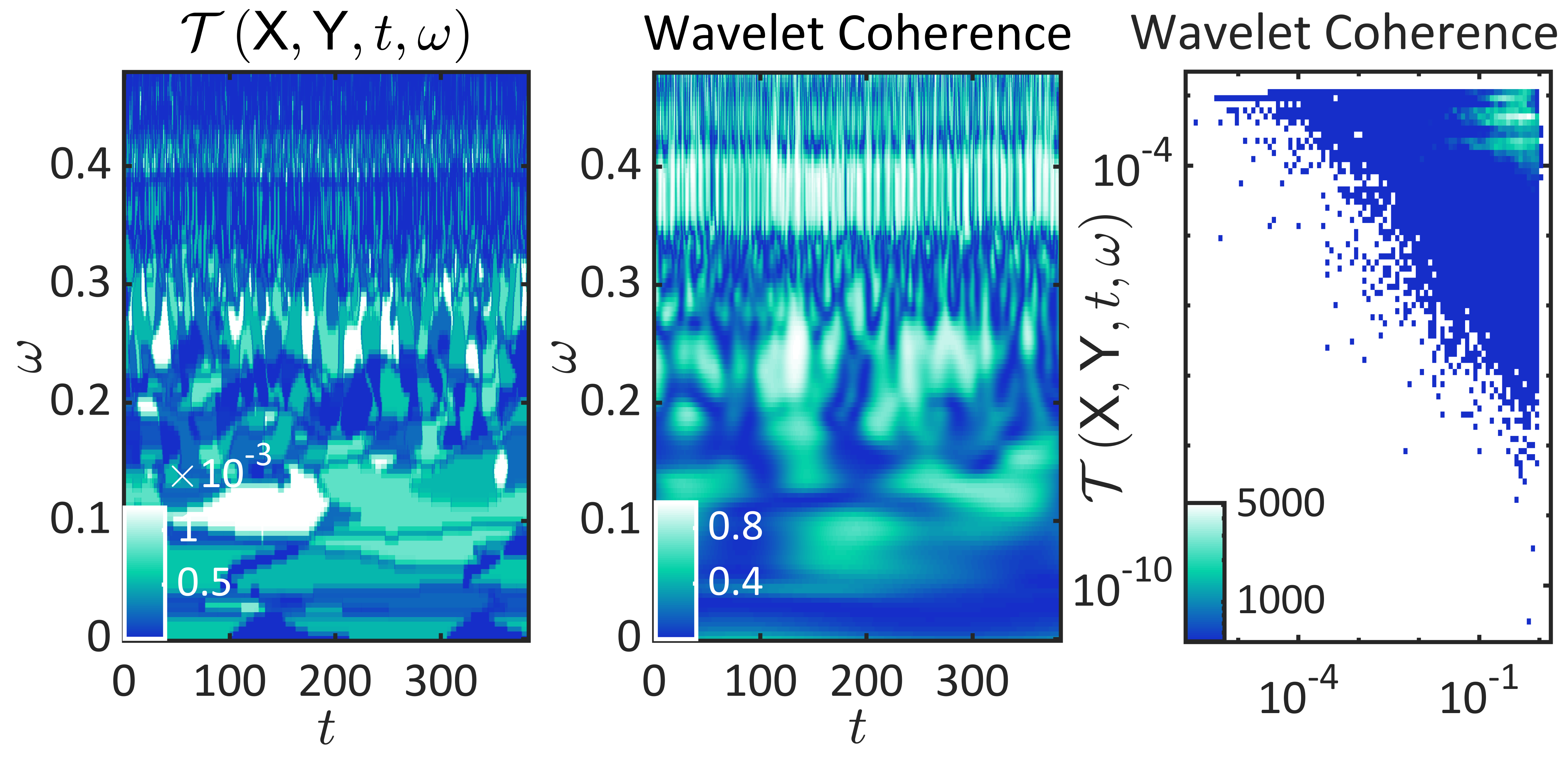}
     \caption{\label{G1c}}
     \end{subfigure}
      \hfill
     \begin{subfigure}[b]{0.48\columnwidth}
         \includegraphics[width=\columnwidth]{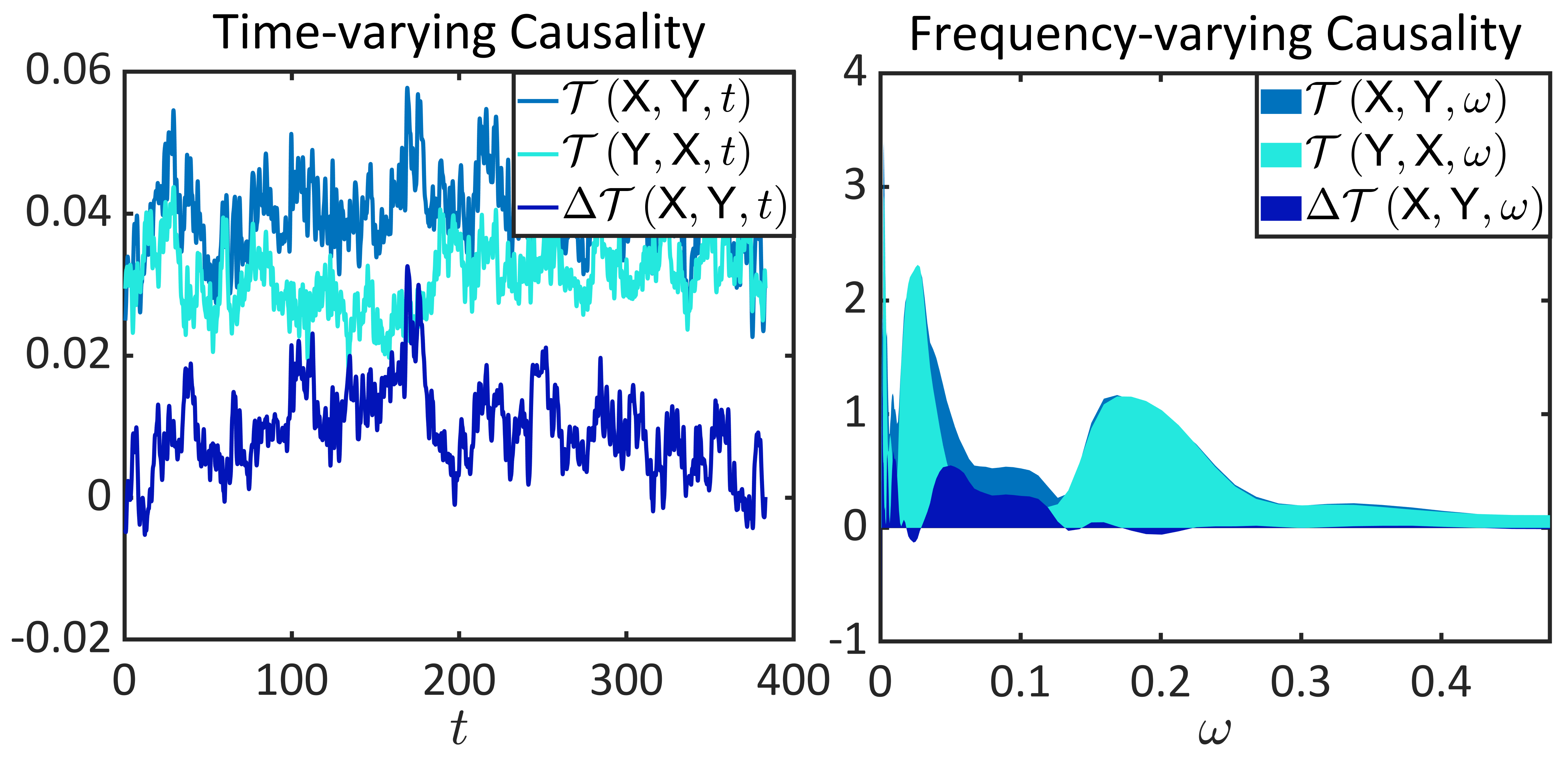}
         \caption{\label{G1d}}
     \end{subfigure}
      \hfill
        \caption{Fourier-domain transfer entropy spectrum analysis. \textbf{(a)} The Fourier-domain representation of the NIRS data. Please note that the frequency $\omega$ is given as cycles per sample (normalized frequency) in our paper. \textbf{(b)} A coarse-grained symbolization with $\left(\lambda,\theta\right)=\left(3,3\right)$ (upper parallel) and a fine-grained symbolization with $\left(\lambda,\theta\right)=\left(50,20\right)$ (bottom parallel). \textbf{(c)} The Fourier-domain transfer entropy spectrum, the wavelet coherence, and their relation. \textbf{(d)} Time-varying and frequency-varying transfer entropy as well as directional information transfer.}
\end{figure}

To overcome the deficiency of probability density estimation (see \textbf{(III)}), we transform these spectra by symbolization. Specifically, we generalize the approach in \cite{bandt2002permutation} to implement a two-dimensional symbolization $\widehat{\mathcal{F}}\left(\mathsf{X},t,\omega\right)$ with a time symbolization order $\lambda$ and a frequency symbolization order $\theta$
     \begin{align}
         \widehat{\mathcal{F}}\left(\mathsf{X},t,\omega\right)&=\big(\Lambda_{\mathsf{X}}\left(t,\omega\right),\Theta_{\mathsf{X}}\left(t,\omega\right)\big),\;\forall t\geq\lambda,\;\omega\geq\theta,\label{EQ5}\\
         \Lambda_{\mathsf{X}}\left(t,\omega\right)&=\sum_{i=1}^{\lambda}\left(\Gamma_{\Lambda}\left(i\right)-1\right)\lambda^{\lambda-i},\label{EQ6}\\
         \Theta_{\mathsf{X}}\left(t,\omega\right)&=\sum_{i=1}^{\theta}\left(\Gamma_{\Theta}\left(i\right)-1\right)\theta^{\theta-i},\label{EQ7}\\
         \Gamma_{\Lambda}&=\gamma\big(\mathcal{F}\left(\mathsf{X},t-\lambda,\omega\right),\ldots,\mathcal{F}\left(\mathsf{X},t,\omega\right)\big),\label{EQ8}\\
         \Gamma_{\Theta}&=\gamma\big(\mathcal{F}\left(\mathsf{X},t,\omega-\theta\right),\ldots,\mathcal{F}\left(\mathsf{X},t,\omega\right)\big).\label{EQ9}
     \end{align}
     The above formulas map each $\mathcal{F}\left(\mathsf{X},t,\omega\right)$ to a two-dimensional coordinate $\big(\Lambda_{\mathsf{X}}\left(t,\omega\right),\Theta_{\mathsf{X}}\left(t,\omega\right)\big)$. The first coordinate component $\Lambda_{\mathsf{X}}\left(t,\omega\right)$ corresponds to the symbolization along the time-line while the second one corresponds to the symbolization along the frequency-line. Both coordinate components are decimal numbers. The symbolization relays on a permutation $\gamma\left(\cdot\right)$ that returns the order of a sequence in ascending sort (e.g., $\gamma\left(1,6,3,10\right)=\left(1,3,2,4\right)$ because $1<3<6<10$). If the order minus one (e.g., $\left(1,3,2,4\right)$ becomes $\left(0,2,1,3\right)$), then it becomes a number in $\lambda$-base (or $\theta$-base) and can be transformed to a decimal number following (\ref{EQ6}-\ref{EQ7}). In Fig. \ref{G1b}, we show the symbolization results of the NIRS data under two different conditions.
     
     The symbolization benefits to our research by embedding $\mathcal{F}\left(\mathsf{X},t,\omega\right)$, a space with large cardinal number (there are a mass of possibilities of the spectrum value of $\mathcal{F}\left(\mathsf{X},t,\omega\right)$ because the power in (\ref{EQ3}) is a continuous variable), to a space with much smaller cardinal number ($\widehat{\mathcal{F}}\left(\mathsf{X},t,\omega\right)$ is a discrete space whose cardinal number is no more than $\lambda!$) \cite{bandt2002permutation}. It functions as a coarse-graining approach to control noises and supports a more adequate sampling of the historical information (e.g., $\mathsf{X}_{\left[t,\beta,\Delta t\right]}$ in (\ref{EQ2})) during probability density estimation (because the variable becomes discrete and its variability is reduced) \cite{porfiri2019transfer}. These benefits should be emphasized especially when $\mathsf{X}$ is non-stationary and a repetitive sampling is not infeasible. In sum, the symbolization in (\ref{EQ5}-\ref{EQ9}) can, to a certain extent, alleviate deficiency \textbf{(III)}. Parameters $\lambda$ and $\theta$ control the granularity of coarse-graining (e.g., see Fig. \ref{G1b}). They should not be too large. Otherwise, the symbolization is too fine-grained and unnecessary \cite{xie2019adaptive}.
     
        After symbolization, the Fourier-domain transfer entropy spectrum can be defined as (\ref{EQ10}), in which we adopt the notions of (\ref{EQ1}) to define an abbreviative representation $\widehat{\mathcal{F}}\left(\mathsf{X}_{\left[t,\beta,\Delta t\right]},\omega\right)=\left(\widehat{\mathcal{F}}\left(\mathsf{X},t-\beta\Delta t,\omega\right),\ldots,\widehat{\mathcal{F}}\left(\mathsf{X},t,\omega\right)\right)$. Although formula (\ref{EQ10}) is consistent with (\ref{EQ2}), it does not average or sum transfer entropy across time and frequency for a single value. Instead, the transfer entropy in (\ref{EQ10}) becomes a function of time and frequency, partly overcoming deficiencies \textbf{(I-II)}. Note that $\mathcal{T}\left(\mathsf{X},\mathsf{Y},t,\omega\right)$ in (\ref{EQ10}) can be either positive or negative, respectively corresponding to the case where the added information of $\widehat{\mathcal{F}}\left(\mathsf{X}_{\left[t,\beta,\Delta t\right]},\omega\right)$ reduces or increases the uncertainty of $\widehat{\mathcal{F}}\left(\mathsf{Y},t,\omega\right)$ given $\widehat{\mathcal{F}}\left(\mathsf{Y}_{\left[t,\beta,\Delta t\right]},\omega\right)$ (there is no limitation of the directionality of regulating effects in (\ref{EQ1})). To keep consistency with the Granger causality that identifies potential causes through uncertainty (error) reduction \cite{granger1969investigating}, we refer to $\widehat{\mathcal{F}}\left(\mathsf{X}_{\left[t,\beta,\Delta t\right]},\omega\right)$ as a cause of $\widehat{\mathcal{F}}\left(\mathsf{Y},t,\omega\right)$ when $\widehat{\mathcal{F}}\left(\mathsf{X}_{\left[t,\beta,\Delta t\right]},\omega\right)$ is positive. A negative $\widehat{\mathcal{F}}\left(\mathsf{X}_{\left[t,\beta,\Delta t\right]},\omega\right)$ corresponds to a decoupling effect on $\widehat{\mathcal{F}}\left(\mathsf{Y},t,\omega\right)$ and $\widehat{\mathcal{F}}\left(\mathsf{X}_{\left[t,\beta,\Delta t\right]},\omega\right)$ provided by $\widehat{\mathcal{F}}\left(\mathsf{X}_{\left[t,\beta,\Delta t\right]},\omega\right)$.

    \begin{align}
 \mathcal{T}\left(\mathsf{X},\mathsf{Y},t,\omega\right)=&\mathcal{P}\left(\widehat{\mathcal{F}}\left(\mathsf{Y},t,\omega\right),\widehat{\mathcal{F}}\left(\mathsf{X}_{\left[t,\beta,\Delta t\right]},\omega\right),\widehat{\mathcal{F}}\left(\mathsf{Y}_{\left[t,\beta,\Delta t\right]},\omega\right)\right)\notag\\&\log\frac{\mathcal{P}\left(\widehat{\mathcal{F}}\left(\mathsf{Y},t,\omega\right),\widehat{\mathcal{F}}\left(\mathsf{X}_{\left[t,\beta,\Delta t\right]},\omega\right)\mid\widehat{\mathcal{F}}\left(\mathsf{Y}_{\left[t,\beta,\Delta t\right]},\omega\right)\right)}{\mathcal{P}\left(\widehat{\mathcal{F}}\left(\mathsf{Y},t,\omega\right)\mid\widehat{\mathcal{F}}\left(\mathsf{Y}_{\left[t,\beta,\Delta t\right]},\omega\right)\right)\mathcal{P}\left(\widehat{\mathcal{F}}\left(\mathsf{X}_{\left[t,\beta,\Delta t\right]},\omega\right)\mid\widehat{\mathcal{F}}\left(\mathsf{Y}_{\left[t,\beta,\Delta t\right]},\omega\right)\right)}.\label{EQ10}
 \end{align}
 
 In Fig. \ref{G1c}, we show the Fourier-domain transfer entropy spectrum of the NIRS data. The symbolization is implemented under a relatively coarse-grained condition $\left(\lambda,\theta\right)=\left(3,3\right)$. Each $\mathcal{T}\left(\mathsf{X},\mathsf{Y},t,\omega\right)$ is measured with $\beta=5$. For convenience, we set each negative $\mathcal{T}\left(\mathsf{X},\mathsf{Y},t,\omega\right)$ as $0$ (this setting can be excluded when the decoupling effects are of interests). In our results, we can see relatively strong and oscillating causality between $\mathsf{X}$ and $\mathsf{Y}$ around $\omega\sim0.1$, $\omega\sim0.25$, and $\omega\sim0.4$. The causality in other frequency regions is relatively weak. We also compare $\mathcal{T}\left(\mathsf{X},\mathsf{Y},t,\omega\right)$ with the well-known wavelet coherence spectrum \cite{torrence1998practical}, which quantifies the time-varying and linear coherence between $\mathsf{X}$ and $\mathsf{Y}$. Consistent with the common knowledge, the causality measured by $\mathcal{T}\left(\mathsf{X},\mathsf{Y},t,\omega\right)$ can not be trivially simplified as linear coherence because these two metrics have no clear accordant pattern.  
 
 \begin{figure}
     \centering
     \begin{subfigure}[b]{0.48\columnwidth}
         \includegraphics[width=\columnwidth]{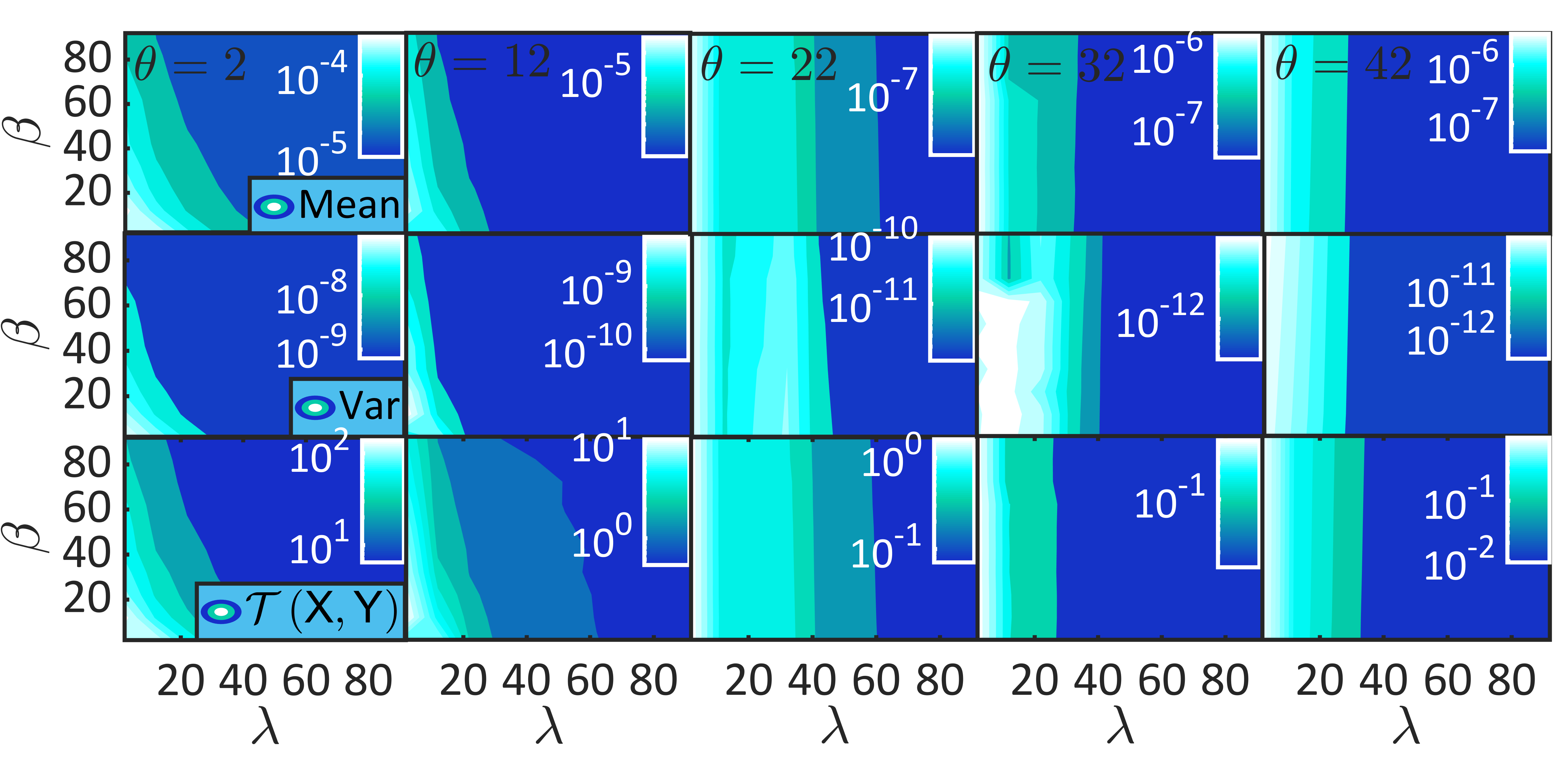}
        \caption{\label{G2a}}
     \end{subfigure}
     \hfill
     \begin{subfigure}[b]{0.48\columnwidth}
         \includegraphics[width=\columnwidth]{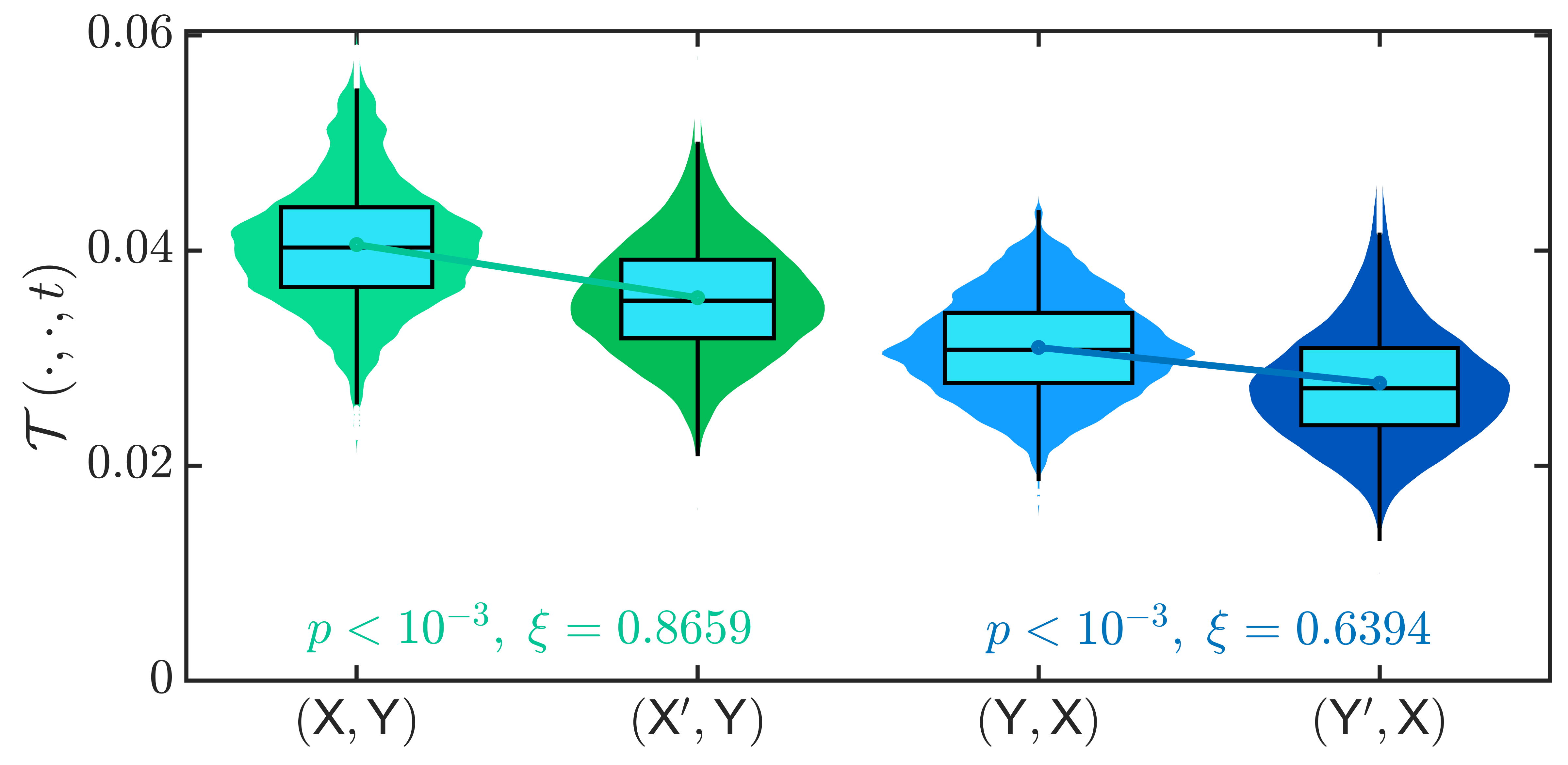}
     \caption{\label{G2b}}
     \end{subfigure}
     \hfill
     \begin{subfigure}[b]{0.48\columnwidth}
         \includegraphics[width=\columnwidth]{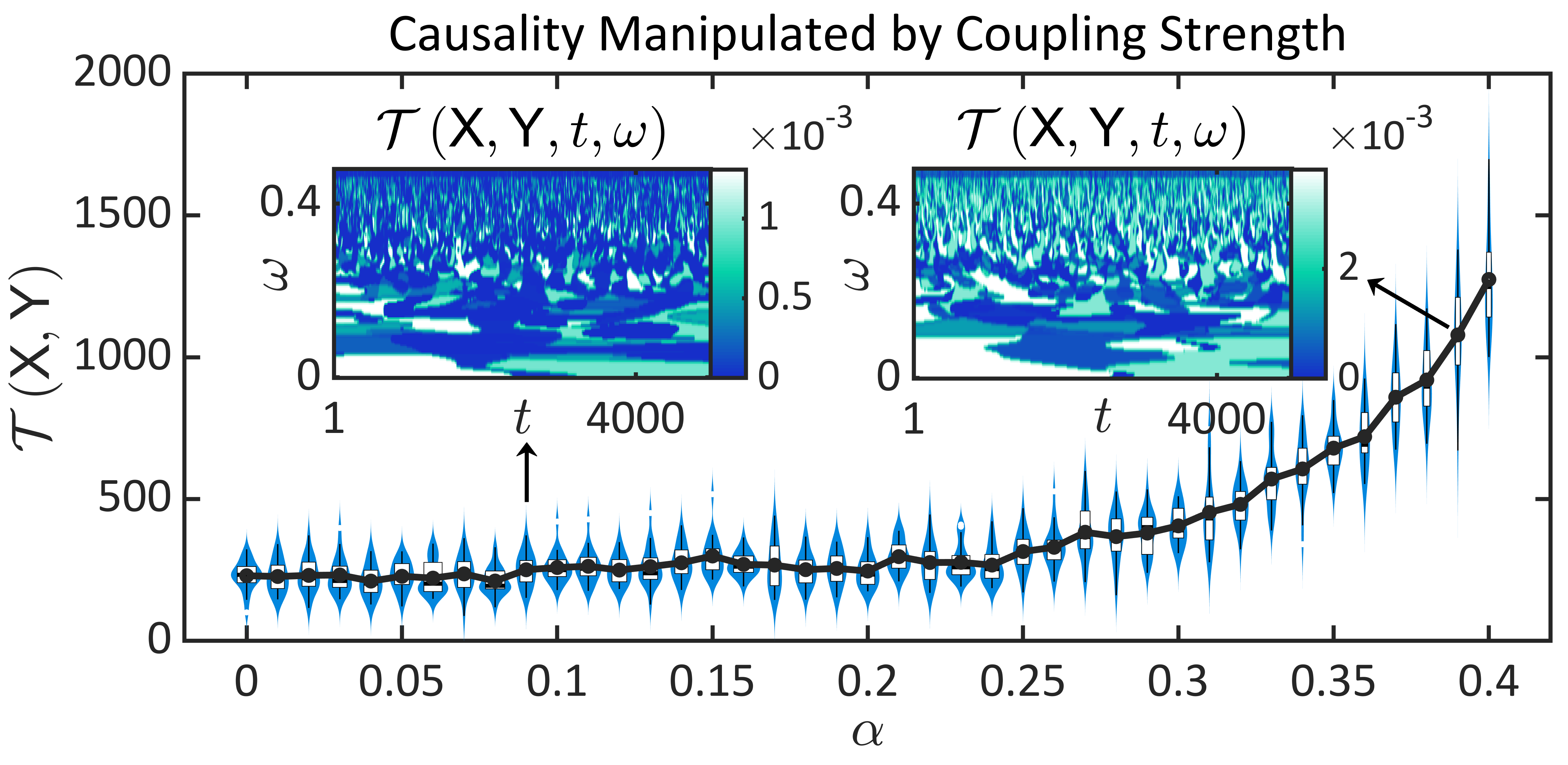}
     \caption{\label{G2c}}
     \end{subfigure}
      \hfill
     \begin{subfigure}[b]{0.48\columnwidth}
         \includegraphics[width=\columnwidth]{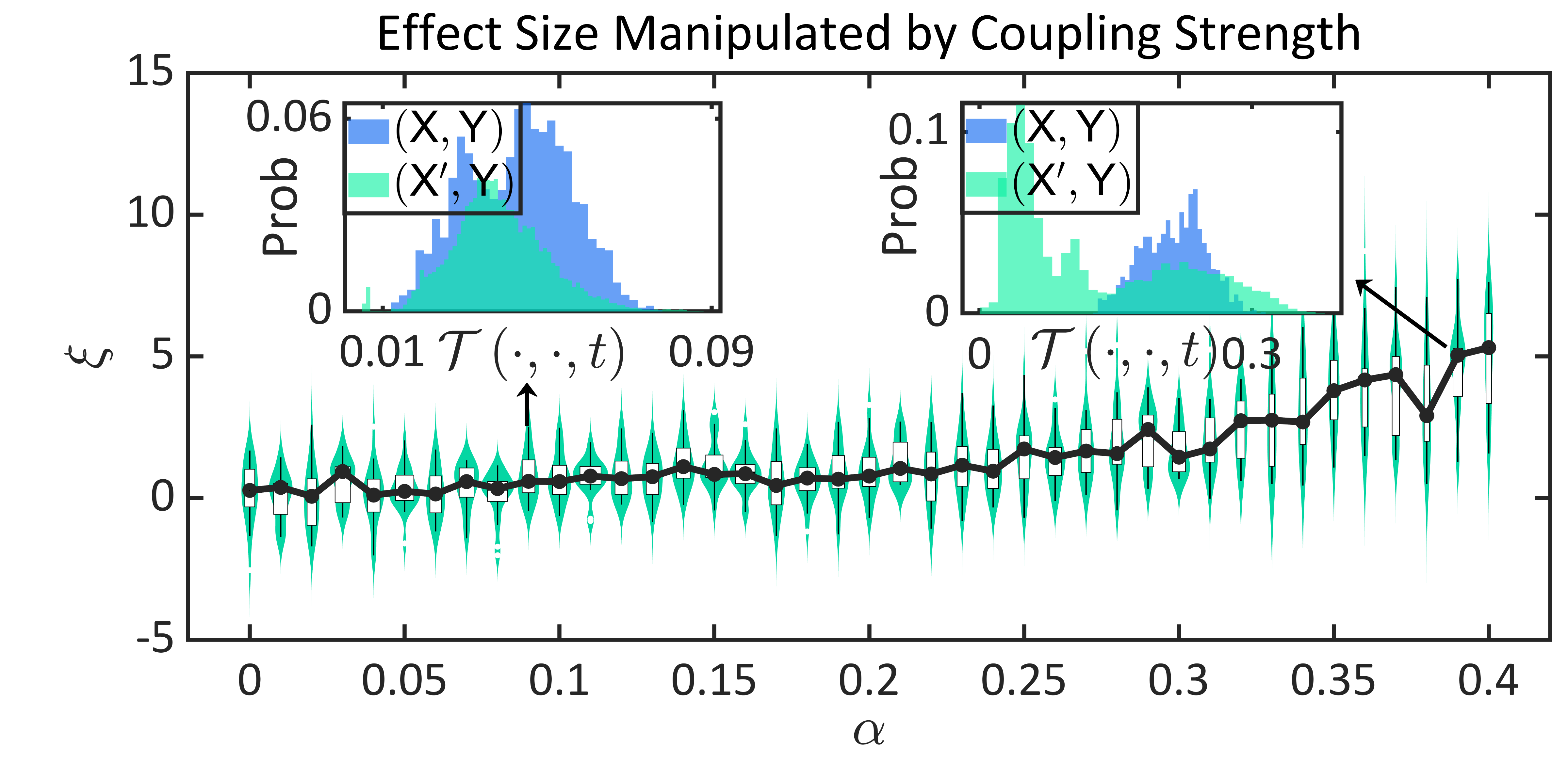}
         \caption{\label{G2d}}
     \end{subfigure}
      \hfill
        \caption{Validity analysis of the Fourier-domain transfer entropy spectrum. \textbf{(a)} The dependence of parameter selections. \textbf{(b)} Statistic significance analysis of the Fourier-domain transfer entropy spectrum in Fig. \ref{G1c}-\ref{G1d}. \textbf{(c-d)} The Fourier-domain transfer entropy spectrum and its effect size is manipulated by the coupling strength $\alpha$.}
\end{figure}
     
     Meanwhile, we can sum $\mathcal{T}\left(\mathsf{X},\mathsf{Y},t,\omega\right)$ over frequency at each moment (see (\ref{EQ11})), over time at each frequency (see (\ref{EQ12})), or over both time and frequency (see (\ref{EQ13})). 
\begin{align}
\mathcal{T}\left(\mathsf{X},\mathsf{Y},t\right)&=\sum_{\omega}\mathcal{T}\left(\mathsf{X},\mathsf{Y},t,\omega\right),\;\forall t,\label{EQ11}\\
\mathcal{T}\left(\mathsf{X},\mathsf{Y},\omega\right)&=\sum_{t}\mathcal{T}\left(\mathsf{X},\mathsf{Y},t,\omega\right),\;\forall \omega,\label{EQ12}\\
\mathcal{T}\left(\mathsf{X},\mathsf{Y}\right)&=\sum_{t,\omega}\mathcal{T}\left(\mathsf{X},\mathsf{Y},t,\omega\right).\label{EQ13}
\end{align}
Formulas (\ref{EQ11}-\ref{EQ13}) denote the varying Fourier-domain transfer entropy across time, frequency, and all possibilities, respectively. Note that $\mathcal{T}\left(\mathsf{X},\mathsf{Y}\right)$ is always non-negative. Combining (\ref{EQ11}-\ref{EQ13}) with the asymmetry property of transfer entropy, we define
\begin{align}
    \Delta\mathcal{T}\left(\mathsf{X},\mathsf{Y},t\right)&=\mathcal{T}\left(\mathsf{X},\mathsf{Y},t\right)-\mathcal{T}\left(\mathsf{Y},\mathsf{X},t\right),\;\forall t,\label{EQ14}\\
    \Delta\mathcal{T}\left(\mathsf{X},\mathsf{Y},\omega\right)&=\mathcal{T}\left(\mathsf{X},\mathsf{Y},\omega\right)-\mathcal{T}\left(\mathsf{Y},\mathsf{X},\omega\right),\;\forall \omega,\label{EQ15}\\
\Delta\mathcal{T}\left(\mathsf{X},\mathsf{Y}\right)&=\mathcal{T}\left(\mathsf{X},\mathsf{Y}\right)-\mathcal{T}\left(\mathsf{Y},\mathsf{X}\right)\label{EQ16}
\end{align}
 to quantify the directional information transfer between $\mathsf{X}$ and $\mathsf{Y}$. In Fig. \ref{G1d}, we illustrate these quantities based on the results in Fig. \ref{G1c}. A stronger information transfer from $\mathsf{X}$ to $\mathsf{Y}$ can be observed in both time- and frequency-line, surpassing the transfer from $\mathsf{Y}$ to $\mathsf{X}$. Such a directional transfer is significant when $\omega\sim 0.1$.
 
 \section{Validity of Fourier-domain transfer entropy spectrum}
After proposing the Fourier-domain transfer entropy spectrum, we turn to analyze its validity. 
 
 A valid causality metric is demanded to have clear dependence relations with all involved parameters. Although the Fourier-domain transfer entropy spectrum is a model-free metric and has no pre-assumption on system properties, the symbolization and the transfer entropy definition itself still introduce three parameters: $\lambda$, $\theta$, and $\beta$. Serving as coarse-graining parameters, $\lambda$ and $\theta$ determine the granularity of symbolization. Sufficiently large quantities of $\lambda$ and $\theta$ may not only make symbolization meaningless \cite{xie2019adaptive} but also cause obstacles in probability density estimation (because the variability of variables increase). An in-exhaustive sampling frequently underestimates specific parts of the probability density, especially in high-dimensional spaces. Therefore, we speculate that $\mathcal{T}\left(\mathsf{X},\mathsf{Y},t,\omega\right)$ may be underestimated when $\lambda$ and $\theta$ become too large. In Fig. \ref{G2a}, we calculate $\mathcal{T}\left(\mathsf{X},\mathsf{Y},t,\omega\right)$ in the NIRS data with different parameters. For convenience, we set each negative $\mathcal{T}\left(\mathsf{X},\mathsf{Y},t,\omega\right)$ as $0$. Consistent with our inference, quantities $\mathbb{E}_{t,\omega}\mathcal{T}\left(\mathsf{X},\mathsf{Y},t,\omega\right)$ (expectation), $\operatorname*{Var}_{t,\omega}\mathcal{T}\left(\mathsf{X},\mathsf{Y},t,\omega\right)$ (variance), and $\mathcal{T}\left(\mathsf{X},\mathsf{Y}\right)$ approximate to $0$ as $\lambda$ and $\theta$ increase. Similar effects can be observed on the lag parameter $\beta$ but become much more slight. In sum, the Fourier-domain transfer entropy spectrum is expected to be implemented with relatively small $\lambda$ and $\theta$. The selection of $\beta$ is less limited and can be carried out according to different demands.   
 
 Moreover, a valid causality metric should be equipped with a statistic significance test. Information-theoretical metrics, including our Fourier-domain transfer entropy spectrum, frequently involve biases in finite data sets \cite{panzeri2007correcting}. Therefore, the absolute value of $\mathcal{T}\left(\mathsf{X},\mathsf{Y},t,\omega\right)$ only has limited meaning until being statistically tested. Combining the time-shifting surrogates \cite{andrzejak2006detecting,wagner2010detection,martini2011inferring} with the permutation test \cite{phipson2010permutation,maris2007nonparametric}, we design the significance test as following: \textbf{(1)} Generate a set of surrogates $\{\mathsf{X}^{\prime}_{\delta}\vert\mathsf{X}^{\prime}_{\delta}\left(t\right)=\mathsf{X}\left(t+\delta\right),\;\delta\in\mathbb{N}^{+}\}$ and calculate $\{\mathcal{T}\left(\mathsf{X}^{\prime}_{\delta},\mathsf{Y},t\right)\}_{t}$ for each surrogate $\mathsf{X}^{\prime}_{\delta}$; \textbf{(2)} Implement permutation test for the difference in mean value between $\{\mathcal{T}\left(\mathsf{X},\mathsf{Y},t\right)\}_{t}$ and $\bigcup_{\delta}\{\mathcal{T}\left(\mathsf{X}^{\prime}_{\delta},\mathsf{Y},t\right)\}_{t}$ to calculate the statistic significance $p$ and the effect size $\xi$. A similar idea can be seen in \cite{lindner2011trentool}. In practice, a weakest condition with $p\leq 10^{-2}$ and $\xi\geq 0.2$ (in the comparison of mean values, $\xi\sim0.2$, $\xi\sim0.5$, and $\xi\sim0.8$ correspond to small, medium, and high effect sizes, respectively) should be satisfied if the Fourier-domain transfer entropy spectrum is statistically significant. In Fig. \ref{G2b}, we test the results in Fig. \ref{G1c}-\ref{G1d}, where $50$ surrogates are generated by randomizing $\delta\in\left[3,100\right]$. The generated $\bigcup_{\delta}\{\mathcal{T}\left(\mathsf{X}^{\prime}_{\delta},\mathsf{Y},t\right)\}_{t}$ features a large sample size of $191350$, supporting a robust statistic test. We implement a large scale permutation test with $5000$ permutations, demonstrating the statistic significance of $\mathcal{T}\left(\mathsf{X},\mathsf{Y},t\right)$ and $\mathcal{T}\left(\mathsf{Y},\mathsf{X},t\right)$ with ideal effect sizes. Meanwhile, the effect size on $\mathcal{T}\left(\mathsf{Y},\mathsf{X},t\right)$ is smaller than that on $\mathcal{T}\left(\mathsf{X},\mathsf{Y},t\right)$, corroborating our observations in Fig. \ref{G1d} about the directional information transfer in the NIRS data.
 
Finally, a valid causality metric is expected to be sensitive to the changes of coupling strength between $\mathsf{X}$ and $\mathsf{Y}$. In other words, $\mathcal{T}\left(\mathsf{X},\mathsf{Y},t,\omega\right)$ should be regulated as consequences if we manipulate the coupling strength. To realize the manipulation, we consider two diffusively coupled logistic oscillators $\mathsf{X}\rightarrow\mathsf{Y}$ in Fig. \ref{G2c}-\ref{G2d}. Notion $\rightarrow$ denotes the direction of coupling. Oscillators $\mathsf{X}$ and $\mathsf{Y}$ are defined with a bifurcation parameter $\mu=4$, and their coupling strength is denoted by $\alpha$. For more mathematical details, we refer to \cite{lloyd1995coupled}. We manipulate the coupling strength $\alpha$ from $0$ to $0.4$ and apply an open-source algorithm \cite{cami} to generate the time series of $\mathsf{X}$ and $\mathsf{Y}$ under each $\alpha$ condition. The generation is repeated $50$ times in each case, corresponding to different random initial values. For each pair of time series, we calculate the Fourier-domain transfer entropy spectrum with $\lambda=\theta=3$ and $\beta=1$. Meanwhile, $5$ surrogates are generated by randomizing $\delta\in\left[3,100\right]$, leading to a large sample size $24985$ in the statistic significance test. The number of permutations is set as $5000$. In Fig. \ref{G2c}-\ref{G2d}, we demonstrate that the Fourier-domain transfer entropy spectrum and its effect sizes (with $p<10^{-2}$) increase with the coupling strength $\alpha$. In other words, the quantified latent causality increases as a consequence of coupling strength enhancement.

 \section{Conclusion}
 We present the Fourier-domain transfer entropy spectrum, a model-free metric of causality, to extract the time-varying latent causality among different system components of an arbitrary pair of systems in the Fourier-domain. Built on the wavelet transform and symbolization, our approach can partly alleviate deficiencies \textbf{(I-III)}, generally applicable in various non-stationary or high-dimensional processes (e.g., neuroscience data). We have demonstrated the validity of our approach in the aspects of parameter dependence, statistic significance test, and sensibility. An open-source release of our algorithm can be seen in \cite{yang2021tesinfd}.
 
\section{Acknowledgements}
This project is supported by the Artificial and General Intelligence Research Program of Guo Qiang Research Institute at Tsinghua University (2020GQG1017) as well as the Tsinghua University Initiative Scientific Research Program. Correspondence of this research should be addressed to Y.Y.W, Z.Y.Z, and P.S.

\bibliographystyle{plain}
\bibliography{references}  

\begin{thebibliography}{64}
\providecommand{\natexlab}[1]{#1}
\providecommand{\url}[1]{\texttt{#1}}
\expandafter\ifx\csname urlstyle\endcsname\relax
  \providecommand{\doi}[1]{doi: #1}\else
  \providecommand{\doi}{doi: \begingroup \urlstyle{rm}\Url}\fi

\bibitem[Pikovsky et~al.(2003)Pikovsky, Kurths, Rosenblum, and
  Kurths]{pikovsky2003synchronization}
Arkady Pikovsky, Jurgen Kurths, Michael Rosenblum, and J{\"u}rgen Kurths.
\newblock \emph{Synchronization: a universal concept in nonlinear sciences}.
\newblock Number~12. Cambridge university press, 2003.

\bibitem[Pereda et~al.(2005)Pereda, Quiroga, and
  Bhattacharya]{pereda2005nonlinear}
Ernesto Pereda, Rodrigo~Quian Quiroga, and Joydeep Bhattacharya.
\newblock Nonlinear multivariate analysis of neurophysiological signals.
\newblock \emph{Progress in neurobiology}, 77\penalty0 (1-2):\penalty0 1--37,
  2005.

\bibitem[Hlav{\'a}{\v{c}}kov{\'a}-Schindler
  et~al.(2007)Hlav{\'a}{\v{c}}kov{\'a}-Schindler, Palu{\v{s}}, Vejmelka, and
  Bhattacharya]{hlavavckova2007causality}
Katerina Hlav{\'a}{\v{c}}kov{\'a}-Schindler, Milan Palu{\v{s}}, Martin
  Vejmelka, and Joydeep Bhattacharya.
\newblock Causality detection based on information-theoretic approaches in time
  series analysis.
\newblock \emph{Physics Reports}, 441\penalty0 (1):\penalty0 1--46, 2007.

\bibitem[Granger(1969)]{granger1969investigating}
Clive~WJ Granger.
\newblock Investigating causal relations by econometric models and
  cross-spectral methods.
\newblock \emph{Econometrica: journal of the Econometric Society}, pages
  424--438, 1969.

\bibitem[Marinazzo et~al.(2008)Marinazzo, Pellicoro, and
  Stramaglia]{marinazzo2008kernel}
Daniele Marinazzo, Mario Pellicoro, and Sebastiano Stramaglia.
\newblock Kernel method for nonlinear granger causality.
\newblock \emph{Physical review letters}, 100\penalty0 (14):\penalty0 144103,
  2008.

\bibitem[Ancona et~al.(2004)Ancona, Marinazzo, and
  Stramaglia]{ancona2004radial}
Nicola Ancona, Daniele Marinazzo, and Sebastiano Stramaglia.
\newblock Radial basis function approach to nonlinear granger causality of time
  series.
\newblock \emph{Physical Review E}, 70\penalty0 (5):\penalty0 056221, 2004.

\bibitem[Ho et~al.(2004)Ho, Hung, and Jiang]{ho2004phase}
Ming-Chung Ho, Yao-Chen Hung, and I-Min Jiang.
\newblock Phase synchronization in inhomogeneous globally coupled map lattices.
\newblock \emph{Physics Letters A}, 324\penalty0 (5-6):\penalty0 450--457,
  2004.

\bibitem[Arnold et~al.(2007)Arnold, Liu, and Abe]{arnold2007temporal}
Andrew Arnold, Yan Liu, and Naoki Abe.
\newblock Temporal causal modeling with graphical granger methods.
\newblock In \emph{Proceedings of the 13th ACM SIGKDD international conference
  on Knowledge discovery and data mining}, pages 66--75, 2007.

\bibitem[Liu et~al.(2009)Liu, Lafferty, and Wasserman]{liu2009nonparanormal}
Han Liu, John Lafferty, and Larry Wasserman.
\newblock The nonparanormal: Semiparametric estimation of high dimensional
  undirected graphs.
\newblock \emph{Journal of Machine Learning Research}, 10\penalty0 (10), 2009.

\bibitem[Dhamala et~al.(2008)Dhamala, Rangarajan, and
  Ding]{dhamala2008estimating}
Mukeshwar Dhamala, Govindan Rangarajan, and Mingzhou Ding.
\newblock Estimating granger causality from fourier and wavelet transforms of
  time series data.
\newblock \emph{Physical review letters}, 100\penalty0 (1):\penalty0 018701,
  2008.

\bibitem[Schreiber(2000)]{schreiber2000measuring}
Thomas Schreiber.
\newblock Measuring information transfer.
\newblock \emph{Physical review letters}, 85\penalty0 (2):\penalty0 461, 2000.

\bibitem[Staniek and Lehnertz(2008)]{staniek2008symbolic}
Matth{\"a}us Staniek and Klaus Lehnertz.
\newblock Symbolic transfer entropy.
\newblock \emph{Physical review letters}, 100\penalty0 (15):\penalty0 158101,
  2008.

\bibitem[Lobier et~al.(2014)Lobier, Siebenh{\"u}hner, Palva, and
  Palva]{lobier2014phase}
Muriel Lobier, Felix Siebenh{\"u}hner, Satu Palva, and J~Matias Palva.
\newblock Phase transfer entropy: a novel phase-based measure for directed
  connectivity in networks coupled by oscillatory interactions.
\newblock \emph{Neuroimage}, 85:\penalty0 853--872, 2014.

\bibitem[Lungarella et~al.(2007)Lungarella, Pitti, and
  Kuniyoshi]{lungarella2007information}
Max Lungarella, Alex Pitti, and Yasuo Kuniyoshi.
\newblock Information transfer at multiple scales.
\newblock \emph{Physical Review E}, 76\penalty0 (5):\penalty0 056117, 2007.

\bibitem[Runge et~al.(2012)Runge, Heitzig, Petoukhov, and
  Kurths]{runge2012escaping}
Jakob Runge, Jobst Heitzig, Vladimir Petoukhov, and J{\"u}rgen Kurths.
\newblock Escaping the curse of dimensionality in estimating multivariate
  transfer entropy.
\newblock \emph{Physical review letters}, 108\penalty0 (25):\penalty0 258701,
  2012.

\bibitem[Wollstadt et~al.(2014)Wollstadt, Mart{\'\i}nez-Zarzuela, Vicente,
  D{\'\i}az-Pernas, and Wibral]{wollstadt2014efficient}
Patricia Wollstadt, Mario Mart{\'\i}nez-Zarzuela, Raul Vicente, Francisco~J
  D{\'\i}az-Pernas, and Michael Wibral.
\newblock Efficient transfer entropy analysis of non-stationary neural time
  series.
\newblock \emph{PloS one}, 9\penalty0 (7):\penalty0 e102833, 2014.

\bibitem[Porfiri and Ruiz~Mar{\'\i}n(2019)]{porfiri2019transfer}
Maurizio Porfiri and Manuel Ruiz~Mar{\'\i}n.
\newblock Transfer entropy on symbolic recurrences.
\newblock \emph{Chaos: An Interdisciplinary Journal of Nonlinear Science},
  29\penalty0 (6):\penalty0 063123, 2019.

\bibitem[Restrepo et~al.(2020)Restrepo, Mateos, and
  Schlotthauer]{restrepo2020transfer}
Juan~F Restrepo, Diego~M Mateos, and Gast{\'o}n Schlotthauer.
\newblock Transfer entropy rate through lempel-ziv complexity.
\newblock \emph{Physical Review E}, 101\penalty0 (5):\penalty0 052117, 2020.

\bibitem[Zhang et~al.(2019)Zhang, Simeone, Cvetkovic, Abela, and
  Richardson]{zhang2019itene}
Jingjing Zhang, Osvaldo Simeone, Zoran Cvetkovic, Eugenio Abela, and Mark
  Richardson.
\newblock Itene: Intrinsic transfer entropy neural estimator.
\newblock \emph{arXiv preprint arXiv:1912.07277}, 2019.

\bibitem[Silini and Masoller(2021)]{silini2021fast}
Riccardo Silini and Cristina Masoller.
\newblock Fast and effective pseudo transfer entropy for bivariate data-driven
  causal inference.
\newblock \emph{Scientific Reports}, 11\penalty0 (1):\penalty0 1--13, 2021.

\bibitem[Pearl et~al.(2000)]{pearl2000models}
Judea Pearl et~al.
\newblock Models, reasoning and inference.
\newblock \emph{Cambridge, UK: CambridgeUniversityPress}, 19, 2000.

\bibitem[Diks and Panchenko(2005)]{diks2005note}
Cees Diks and Valentyn Panchenko.
\newblock A note on the hiemstra-jones test for granger non-causality.
\newblock \emph{Studies in nonlinear dynamics \& econometrics}, 9\penalty0 (2),
  2005.

\bibitem[Diks and DeGoede(2001)]{diks2001general}
Cees Diks and Jacob DeGoede.
\newblock A general nonparametric bootstrap test for granger causality.
\newblock \emph{Global analysis of dynamical systems}, pages 391--403, 2001.

\bibitem[Vicente et~al.(2011)Vicente, Wibral, Lindner, and
  Pipa]{vicente2011transfer}
Raul Vicente, Michael Wibral, Michael Lindner, and Gordon Pipa.
\newblock Transfer entropy—a model-free measure of effective connectivity for
  the neurosciences.
\newblock \emph{Journal of computational neuroscience}, 30\penalty0
  (1):\penalty0 45--67, 2011.

\bibitem[Victor(2002)]{victor2002binless}
Jonathan~D Victor.
\newblock Binless strategies for estimation of information from neural data.
\newblock \emph{Physical Review E}, 66\penalty0 (5):\penalty0 051903, 2002.

\bibitem[Vakorin et~al.(2010)Vakorin, Kovacevic, and
  McIntosh]{vakorin2010exploring}
Vasily~A Vakorin, Natasa Kovacevic, and Anthony~R McIntosh.
\newblock Exploring transient transfer entropy based on a group-wise ica
  decomposition of eeg data.
\newblock \emph{Neuroimage}, 49\penalty0 (2):\penalty0 1593--1600, 2010.

\bibitem[Spinney et~al.(2017)Spinney, Prokopenko, and
  Lizier]{spinney2017transfer}
Richard~E Spinney, Mikhail Prokopenko, and Joseph~T Lizier.
\newblock Transfer entropy in continuous time, with applications to jump and
  neural spiking processes.
\newblock \emph{Physical Review E}, 95\penalty0 (3):\penalty0 032319, 2017.

\bibitem[Ursino et~al.(2020)Ursino, Ricci, and Magosso]{ursino2020transfer}
Mauro Ursino, Giulia Ricci, and Elisa Magosso.
\newblock Transfer entropy as a measure of brain connectivity: a critical
  analysis with the help of neural mass models.
\newblock \emph{Frontiers in computational neuroscience}, 14:\penalty0 45,
  2020.

\bibitem[Dimpfl and Peter(2013)]{dimpfl2013using}
Thomas Dimpfl and Franziska~Julia Peter.
\newblock Using transfer entropy to measure information flows between financial
  markets.
\newblock \emph{Studies in Nonlinear Dynamics and Econometrics}, 17\penalty0
  (1):\penalty0 85--102, 2013.

\bibitem[Papana et~al.(2016)Papana, Kyrtsou, Kugiumtzis, and
  Diks]{papana2016detecting}
Angeliki Papana, Catherine Kyrtsou, Dimitris Kugiumtzis, and Cees Diks.
\newblock Detecting causality in non-stationary time series using partial
  symbolic transfer entropy: Evidence in financial data.
\newblock \emph{Computational economics}, 47\penalty0 (3):\penalty0 341--365,
  2016.

\bibitem[Toriumi and Komura(2018)]{toriumi2018investment}
Fujio Toriumi and Kazuki Komura.
\newblock Investment index construction from information propagation based on
  transfer entropy.
\newblock \emph{Computational Economics}, 51\penalty0 (1):\penalty0 159--172,
  2018.

\bibitem[Camacho et~al.(2021)Camacho, Romeu, and
  Ruiz-Marin]{camacho2021symbolic}
Maximo Camacho, Andres Romeu, and Manuel Ruiz-Marin.
\newblock Symbolic transfer entropy test for causality in longitudinal data.
\newblock \emph{Economic Modelling}, 94:\penalty0 649--661, 2021.

\bibitem[Ji et~al.(2018)Ji, Marfatia, and Gupta]{ji2018information}
Qiang Ji, Hardik Marfatia, and Rangan Gupta.
\newblock Information spillover across international real estate investment
  trusts: Evidence from an entropy-based network analysis.
\newblock \emph{The North American Journal of Economics and Finance},
  46:\penalty0 103--113, 2018.

\bibitem[Cover(1999)]{cover1999elements}
Thomas~M Cover.
\newblock \emph{Elements of information theory}.
\newblock John Wiley \& Sons, 1999.

\bibitem[Smirnov(2013)]{smirnov2013spurious}
Dmitry~A Smirnov.
\newblock Spurious causalities with transfer entropy.
\newblock \emph{Physical Review E}, 87\penalty0 (4):\penalty0 042917, 2013.

\bibitem[Kraskov et~al.(2004)Kraskov, St{\"o}gbauer, and
  Grassberger]{kraskov2004estimating}
Alexander Kraskov, Harald St{\"o}gbauer, and Peter Grassberger.
\newblock Estimating mutual information.
\newblock \emph{Physical review E}, 69\penalty0 (6):\penalty0 066138, 2004.

\bibitem[Gardner et~al.(2006)Gardner, Napolitano, and
  Paura]{gardner2006cyclostationarity}
William~A Gardner, Antonio Napolitano, and Luigi Paura.
\newblock Cyclostationarity: Half a century of research.
\newblock \emph{Signal processing}, 86\penalty0 (4):\penalty0 639--697, 2006.

\bibitem[Marks~II(2009)]{marks2009handbook}
Robert~J Marks~II.
\newblock \emph{Handbook of Fourier analysis \& its applications}.
\newblock Oxford University Press, 2009.

\bibitem[Br{\'e}maud(2013)]{bremaud2013mathematical}
Pierre Br{\'e}maud.
\newblock \emph{Mathematical principles of signal processing: Fourier and
  wavelet analysis}.
\newblock Springer Science \& Business Media, 2013.

\bibitem[Bruns(2004)]{bruns2004fourier}
Andreas Bruns.
\newblock Fourier-, hilbert-and wavelet-based signal analysis: are they really
  different approaches?
\newblock \emph{Journal of neuroscience methods}, 137\penalty0 (2):\penalty0
  321--332, 2004.

\bibitem[Rabiner and Gold(1975)]{rabiner1975theory}
Lawrence~R Rabiner and Bernard Gold.
\newblock Theory and application of digital signal processing.
\newblock \emph{Englewood Cliffs: Prentice-Hall}, 1975.

\bibitem[Rioul and Vetterli(1991)]{rioul1991wavelets}
Olivier Rioul and Martin Vetterli.
\newblock Wavelets and signal processing.
\newblock \emph{IEEE signal processing magazine}, 8\penalty0 (4):\penalty0
  14--38, 1991.

\bibitem[Huang(2014)]{huang2014hilbert}
Norden~Eh Huang.
\newblock \emph{Hilbert-Huang transform and its applications}, volume~16.
\newblock World Scientific, 2014.

\bibitem[Kaiser and Hudgins(1994)]{kaiser1994friendly}
Gerald Kaiser and Lonnie~H Hudgins.
\newblock \emph{A friendly guide to wavelets}, volume 300.
\newblock Springer, 1994.

\bibitem[Folland and Sitaram(1997)]{folland1997uncertainty}
Gerald~B Folland and Alladi Sitaram.
\newblock The uncertainty principle: a mathematical survey.
\newblock \emph{Journal of Fourier analysis and applications}, 3\penalty0
  (3):\penalty0 207--238, 1997.

\bibitem[Peng et~al.(2005)Peng, Peter, and Chu]{peng2005improved}
ZK~Peng, W~Tse Peter, and FL~Chu.
\newblock An improved hilbert--huang transform and its application in vibration
  signal analysis.
\newblock \emph{Journal of sound and vibration}, 286\penalty0 (1-2):\penalty0
  187--205, 2005.

\bibitem[Ogden(1997)]{ogden1997essential}
R~Todd Ogden.
\newblock \emph{Essential wavelets for statistical applications and data
  analysis}.
\newblock Springer, 1997.

\bibitem[Chun-Lin(2010)]{chun2010tutorial}
Liu Chun-Lin.
\newblock A tutorial of the wavelet transform.
\newblock \emph{NTUEE, Taiwan}, 2010.

\bibitem[Coifman et~al.(1992)Coifman, Meyer, and
  Wickerhauser]{coifman1992wavelet}
Ronald~R Coifman, Yves Meyer, and Victor Wickerhauser.
\newblock Wavelet analysis and signal processing.
\newblock In \emph{In Wavelets and their applications}. Citeseer, 1992.

\bibitem[Cui et~al.(2012{\natexlab{a}})Cui, Bryant, and Reiss]{cui2012nirs}
Xu~Cui, Daniel~M Bryant, and Allan~L Reiss.
\newblock Nirs-based hyperscanning reveals increased interpersonal coherence in
  superior frontal cortex during cooperation.
\newblock \emph{Neuroimage}, 59\penalty0 (3):\penalty0 2430--2437,
  2012{\natexlab{a}}.

\bibitem[Cui et~al.(2012{\natexlab{b}})Cui, Bryant, and Reiss]{dataset}
Xu~Cui, Daniel~M Bryant, and Allan~L Reiss.
\newblock The nirs data in a built-in directory of matlab, 2012{\natexlab{b}}.
\newblock URL
  \url{https://ww2.mathworks.cn/help/wavelet/ug/wavelet-coherence-of-brain-dynamics.html}.

\bibitem[Bandt and Pompe(2002)]{bandt2002permutation}
Christoph Bandt and Bernd Pompe.
\newblock Permutation entropy: a natural complexity measure for time series.
\newblock \emph{Physical review letters}, 88\penalty0 (17):\penalty0 174102,
  2002.

\bibitem[Xie et~al.(2019)Xie, Gao, Gao, Lv, and Wang]{xie2019adaptive}
Juntai Xie, Jianmin Gao, Zhiyong Gao, Xiaozhe Lv, and Rongxi Wang.
\newblock Adaptive symbolic transfer entropy and its applications in modeling
  for complex industrial systems.
\newblock \emph{Chaos: An Interdisciplinary Journal of Nonlinear Science},
  29\penalty0 (9):\penalty0 093114, 2019.

\bibitem[Torrence and Compo(1998)]{torrence1998practical}
Christopher Torrence and Gilbert~P Compo.
\newblock A practical guide to wavelet analysis.
\newblock \emph{Bulletin of the American Meteorological society}, 79\penalty0
  (1):\penalty0 61--78, 1998.

\bibitem[Panzeri et~al.(2007)Panzeri, Senatore, Montemurro, and
  Petersen]{panzeri2007correcting}
Stefano Panzeri, Riccardo Senatore, Marcelo~A Montemurro, and Rasmus~S
  Petersen.
\newblock Correcting for the sampling bias problem in spike train information
  measures.
\newblock \emph{Journal of neurophysiology}, 98\penalty0 (3):\penalty0
  1064--1072, 2007.

\bibitem[Andrzejak et~al.(2006)Andrzejak, Ledberg, and
  Deco]{andrzejak2006detecting}
RG~Andrzejak, A~Ledberg, and G~Deco.
\newblock Detecting event-related time-dependent directional couplings.
\newblock \emph{New Journal of Physics}, 8\penalty0 (1):\penalty0 6, 2006.

\bibitem[Wagner et~al.(2010)Wagner, Fell, and Lehnertz]{wagner2010detection}
T~Wagner, J~Fell, and K~Lehnertz.
\newblock The detection of transient directional couplings based on phase
  synchronization.
\newblock \emph{New Journal of Physics}, 12\penalty0 (5):\penalty0 053031,
  2010.

\bibitem[Martini et~al.(2011)Martini, Kranz, Wagner, and
  Lehnertz]{martini2011inferring}
Marcel Martini, Thorsten~A Kranz, Tobias Wagner, and Klaus Lehnertz.
\newblock Inferring directional interactions from transient signals with
  symbolic transfer entropy.
\newblock \emph{Physical review E}, 83\penalty0 (1):\penalty0 011919, 2011.

\bibitem[Phipson and Smyth(2010)]{phipson2010permutation}
Belinda Phipson and Gordon~K Smyth.
\newblock Permutation p-values should never be zero: calculating exact p-values
  when permutations are randomly drawn.
\newblock \emph{Statistical applications in genetics and molecular biology},
  9\penalty0 (1), 2010.

\bibitem[Maris and Oostenveld(2007)]{maris2007nonparametric}
Eric Maris and Robert Oostenveld.
\newblock Nonparametric statistical testing of eeg-and meg-data.
\newblock \emph{Journal of neuroscience methods}, 164\penalty0 (1):\penalty0
  177--190, 2007.

\bibitem[Lindner et~al.(2011)Lindner, Vicente, Priesemann, and
  Wibral]{lindner2011trentool}
Michael Lindner, Raul Vicente, Viola Priesemann, and Michael Wibral.
\newblock Trentool: A matlab open source toolbox to analyse information flow in
  time series data with transfer entropy.
\newblock \emph{BMC neuroscience}, 12\penalty0 (1):\penalty0 1--22, 2011.

\bibitem[Lloyd(1995)]{lloyd1995coupled}
Alun~L Lloyd.
\newblock The coupled logistic map: a simple model for the effects of spatial
  heterogeneity on population dynamics.
\newblock \emph{Journal of Theoretical Biology}, 173\penalty0 (3):\penalty0
  217--230, 1995.

\bibitem[Valencio and Baptista(2018)]{cami}
Arthur Valencio and Murilo da~Silva Baptista.
\newblock Coupled logistic maps: functions for generating the time-series from
  networks of coupled logistic systems, 2018.
\newblock Open source codes for MATLAB. Available at
  \url{https://github.com/artvalencio/coupled-logistic-maps}.

\bibitem[Tian et~al.(2021)Tian, Wang, Zhang, and Sun]{yang2021tesinfd}
Yang Tian, Yaoyuan Wang, Ziyang Zhang, and Pei Sun.
\newblock Transfer-entropy-spectrum-in-the-fourier-domain, 2021.
\newblock Open source codes available at
  \url{https://github.com/doloMing/Transfer-entropy-spectrum-in-the-Fourier-domain}.

\end{thebibliography}






\end{document}